\documentclass[conference,10pt]{IEEEtran}

\usepackage[pdftex]{graphicx}
\usepackage{tikz}
\usepackage{pifont}

\usepackage[english]{babel}
\usepackage[latin1]{inputenc}

\usepackage{amssymb}
\usepackage{url}

\RequirePackage{bm} 
\RequirePackage{amsmath}
\RequirePackage{amsfonts}
\RequirePackage{amssymb}
\RequirePackage{mathrsfs}
\RequirePackage{dsfont} 
\RequirePackage[single]{accents} 
\RequirePackage{exscale} 

\newcommand{\rv}[1]{\ensuremath{{\boldsymbol{#1}}}}
\renewcommand{\vec}[1]{\ensuremath{{\underline{#1}}}}
\newcommand{\rvec}[1]{\ensuremath{{\boldsymbol{\underline{#1}}}}}
\newcommand{\mat}[1]{{\ensuremath{{\mathbf{#1}}}}}

\def\atan2{\operatorname{atan2}}
\renewcommand{\arg}{\operatorname*{arg}}

\newcommand{\dd}{\operatorname{d}\!}
\DeclareMathOperator{\diag}{diag}

\renewcommand{\Im}[1][KeinImagTeil]{%
  \ifthenelse{\equal{#1}{KeinImagTeil}}{
    \ensuremath{\operatorname{Im}}%
  }{
    \ensuremath{\operatorname{Im}\!\left\{#1\right\}}%
  }
}
\renewcommand{\Re}[1][KeinRealTeil]{%
  \ifthenelse{\equal{#1}{KeinRealTeil}}{
    \ensuremath{\operatorname{Re}}%
  }{
    \ensuremath{\operatorname{Re}\!\left\{#1\right\}}%
  }
}

\def\T{^\mathrm{T}} 
\DeclareMathOperator{\trace}{trace}

\renewcommand{\arctan}{\operatorname{arctan}}

\def\({\left(}
\def\){\right)}

\newcommand{\NewR}{\ensuremath{\mathds{R}}}

\newcommand{\PDF}{probability density function}
\newcommand{\Gauss}{{\mathcal{N}}}

\def\DD{Dirac delta distribution}

\def\PDF{probability density function}

\def\Sec#1{Sec.~\ref{#1}}

\def\Fig#1{Fig.~\ref{#1}}
\def\Tab#1{Table~\ref{#1}}

\def\w{\rv w}
\def\x{\rv x}
\def\y{\rv y}

\def\va{\vec a}
\def\vb{\vec b}

\def\ve{\vec e}
\def\vg{\vec g}
\def\vh{\vec h}

\def\vu{\vec u}
\def\vv{\vec v}
\def\vw{\vec w}
\def\vx{\vec x}
\def\vy{\vec y}
\def\vz{\vec z}

\def\vX{\vec X}

\def\rve{\rvec e}
\def\rvv{\rvec v}
\def\rvw{\rvec w}
\def\rvx{\rvec x}
\def\rvy{\rvec y}
\def\rvz{\rvec z}

\def\rvX{\rvec X}

\def\A{\mat A}
\def\B{\mat B}
\def\C{\mat C}
\def\D{\mat D}
\def\G{\mat G}

\def\H{\mat H}
\def\I{\mat I}
\def\K{\mat K}

\def\P{\mat P}
\def\S{\mat S}
\def\V{\mat V}

\def\hx{\hat x}

\def\hz{\hat z}
\def\hvv{\hat{\vec v}}
\def\hvw{\hat{\vec w}}
\def\hvx{\hat{\vec x}}
\def\hvy{\hat{\vec y}}
\def\hvz{\hat{\vec z}}

\def\hvX{\hat{\vec X}}

\usepackage[squaren,thinspace]{SIunits}
\newcommand{\Meter}[2][]{\unit{#2}#1\meter}

\newcommand{\Rad}[2][]{\unit{#2}#1\rad}

\usepackage{theorem}
\theoremheaderfont{\bfseries}
\theoremstyle{plain}
{\theorembodyfont{\itshape} }
{\theorembodyfont{\itshape} }
{\theorembodyfont{\itshape}}
{\theorembodyfont{\sffamily}\newtheorem{example}     {\small Example}}
{\theorembodyfont{\normalfont} }
{\theorembodyfont{\rmfamily}}

\usepackage{mathtools}
\mathtoolsset{showonlyrefs}

\usepackage{cite} 
\usepackage{stfloats}  
                        
\usepackage{flushend}
                        
\begin{document}
\usetikzlibrary{topaths}
\usetikzlibrary{decorations.pathmorphing}
\usetikzlibrary{fit}
\usetikzlibrary{shapes.geometric}

\newcommand{\RA}{\tikz[baseline=0mm]  \draw[ultra thick, -stealth] (0,1.3mm)--(3.5mm,1.3mm); }
\newcommand{\VA}{\tikz[baseline=0mm]  \draw[ultra thick, -stealth] (0,0mm)--(0mm,-3.5mm); }
\newcommand{\VUA}{\tikz[baseline=0mm]  \draw[ultra thick, -stealth] (0,0mm)--(0mm,3.5mm); }
\newcommand{\DA}{\tikz[baseline=0mm]  \draw[ultra thick, stealth-stealth] (0,1.3mm)--(5.5mm,1.3mm); }

\newcommand{\coordcross}[6]{%
	\draw[->, >=latex] (#1,0) -- (#2,0) node[right,below] {#5};
  \draw[->, >=latex] (0,#3) -- (0,#4) node[left] {#6};
}

\tikzstyle{blockdef}=[rectangle, rounded corners, draw=black, thick, text centered]
\tikzstyle{rblock}=[blockdef, inner sep=4pt, text width=1.3cm]
\tikzstyle{dblock}=[shape=diamond, draw=black, thick, text centered, aspect=1.5]
\tikzstyle{cblock}=[blockdef, circle, inner sep=2pt, text width=2mm]
\tikzstyle{ablock}=[thick, -latex, black]
\tikzstyle{aablock}=[ultra thick, latex-latex, black]
\tikzstyle{fblock}=[draw=black, fill]
\tikzstyle{fitblock}=[blockdef, fill=none, dashed, inner sep = 1mm]

\tikzstyle{hlNone}=[anchor=base, rounded corners=3pt]
\tikzstyle{hlFrameNone}=[anchor=base, rounded corners=3pt, thick]
\tikzstyle{hlFrameBlue}=[hlFrameNone, fill=BlockBlue, draw=HilightBlue]
\tikzstyle{hlFrameRed}=[hlFrameNone, fill=BlockRed, draw=HilightRed]
\tikzstyle{hlBlue}=[hlNone, fill=BlockBlue]
\tikzstyle{hlGreen}=[hlNone, fill=BlockGreen]
\tikzstyle{hlRed}=[hlNone, fill=BlockRed]

\title{Adaptive Gaussian Mixture Filter}
\title{Adaptive Gaussian Mixture Filter\\ Based on Statistical Linearization}
\author{\authorblockN{Marco F. Huber}
\authorblockA{Vari\-able Im\-age Ac\-qui\-si\-tion and Pro\-cess\-ing Re\-search Group\\ Fraun\-hofer In\-st\-itute of Op\-tronics, Sys\-tem Tech\-nolo\-gies and Im\-age Ex\-plo\-ita\-tion IOSB\\
Karlsruhe, Germany\\
Email: marco.huber@ieee.org}
}

\maketitle

\selectlanguage{english}

\begin{abstract}
Gaussian mixtures are a common density representation in nonlinear, non-Gaussian Bayesian state estimation. Selecting an appropriate number of Gaussian components, however, is difficult as one has to trade of computational complexity against estimation accuracy. In this paper, an adaptive Gaussian mixture filter based on statistical linearization is proposed. 
Depending on the nonlinearity of the considered estimation problem, this filter dynamically increases the number of components via splitting.
For this purpose, a measure is introduced that allows for quantifying the locally induced linearization error at each Gaussian mixture component. 
The deviation between the nonlinear and
the linearized state space model is evaluated for determining the splitting direction. The proposed approach is not restricted to a specific statistical linearization method. 
Simulations show the superior estimation performance compared to related approaches and common filtering algorithms.
\end{abstract}


%
\IEEEpeerreviewmaketitle
\section{Introduction}
\label{sec:Introduction}
Bayesian state estimation for nonlinear systems requires an efficient approximation for practical applications as closed-form solutions are not available in general. 
A common approximation technique is the discretization of the state space as done in grid filters or particle filters \cite{Arulampalam_ParticleFiltersTutorial}. Theoretically, these techniques facilitate to approach the true statistics of the state with arbitrary accuracy. But they are only applicable to low-dimensional problems since their computational complexity increases exponentially with the dimension of the state space.

A famous exception that exhibits an analytic solution is the linear Gaussian case. Here, the famous Kalman filter provides optimal results in an efficient manner \cite{Kalman1960}. So-called Gaussian filters try to adapted the Kalman filter equations to nonlinear problems by assuming that the density function of the state can be represented by a Gaussian density. The extended Kalman filter \cite{Simon2006} applies first-order Taylor series expansion for linearization. The unscented Kalman filter  \cite{Wan_2000, Julier_IEEE2004} or the Gaussian estimator \cite{IFAC08_Huber} offer higher order accuracy by employing statistical linearization. 
But in general a single Gaussian density is typically not a sufficient representation for the true density function, which may be skew or multimodal. Thanks to their universal approximator property, Gaussian mixtures \cite{Mazya1996} are a much better approach for approximating complex density functions. 
Examples for Gaussian mixture filters applied to nonlinear estimation are in  \cite{Alspach_GaussianSumApproximation, Simandl_IFAC2005}. 

The estimation accuracy of Gaussian mixture filters significantly depend on the number of Gaussian components used. This number is typically defined by the user. In this paper, a novel Gaussian mixture filter is proposed, which adapts the number of components dynamically and on-line. The nonlinear system and measurement models are linearized locally by means of statistical linearization at each component of the Gaussian mixture. The induced linearization error is quantified by means of the linearization error covariance matrix. Based on this error, a novel moment-preserving splitting procedure is proposed for introducing new mixture components. The component causing the highest linearization error is selected, while splitting is performed in direction of the strongest nonlinearity, i.e., the strongest deviation between the nonlinear model and its linearized version. 
Both linearization and splitting are independent of the used statistical linearization method, which makes the proposed filter versatilely applicable. 

The paper is structured as follows: The Bayesian state estimation problem is formulated in the next section. In \Sec{sec:linearization}, a brief introduction in statistical linearization is given. The novel splitting scheme is derived in \Sec{sec:splitting}. Based on this, \Sec{sec:gaussianmixture} describes the complete adaptive Gaussian mixture filter with all major components. Numerical evaluation by means of simulations is part of \Sec{sec:sim}. The paper closes with concluding remarks.


\section{Problem Formulation}
\label{sec:problem}
In this paper, discrete-time nonlinear dynamic systems 
\begin{align}
	\label{eq:sysmodel}
	\rvx_{k+1} &= \va_k(\rvx_k, \vu_k, \rvw_k)~,
	\\
	\label{eq:msrmodel}
	\rvz_k &= \vh_k(\rvx_k, \rvv_k)
\end{align}
are considered. Here, \eqref{eq:sysmodel} is the dynamics model with the known time-variant nonlinear system function $\va_k(\cdot)$, which propagates the system state\footnote{Random vectors are denoted by boldface letters.} $\rvx_k\in\NewR^{n_x}$ at time step $k$ to time step $k+1$, given the current system input $\vu_k\in\NewR^{n_u}$ and the process noise $\rvw_k\in\NewR^{n_w}$. 
The measurement model is given by \eqref{eq:msrmodel}, where $\vh_k(\cdot)$ is the known time-variant nonlinear measurement function, $\rvz_k\in\NewR^{n_z}$ is the measurement vector, and $\rvv_k\in\NewR^{n_v}$ is the measurement noise. Note that an actual measurement value $\vz_k$ is a realization of the random vector $\rvz_k$ in \eqref{eq:msrmodel}. 

Both noise processes $\rvw_k$ and $\rvv_k$ are assumed to be independent and white. The \PDF s of $\rvw_k$ and $\rvv_k$ are denoted by $f_k^w(\vw_k)$ and $f_k^v(\vv_k)$, respectively. It is assumed that these density functions are described via \emph{Gaussian mixtures}
\begin{align}
	\label{eq:gm_noise}
	f_k^w(\vw_k) &= \sum_{i=1}^{L_k^w} \omega_{k,i}^w\cdot\Gauss(\vw_k; \hvw_{k,i}, \C_{k,i}^w)~,
	\\
	\label{eq:gm_msrnoise}
	f_k^v(\vv_k) &= \sum_{i=1}^{L_k^v} \omega_{k,i}^v\cdot\Gauss(\vv_k; \hvv_{k,i}, \C_{k,i}^v)~,
\end{align}
where $L_k^w$, $L_k^v$ are the numbers of mixture components, $\omega_{k,i}^w$, $\omega_{k,i}^v$ are non-negative weights that sum up to one, and $\Gauss(\vw; \hvw, \C^w)$ is a Gaussian density with mean vector $\hvw$ and covariance matrix $\C^w$. The initial density function $f_0^x(\vx_0)$ of the system state at time step $k=0$ is also assumed to be given as a Gaussian mixture.


Estimating the system state from noisy measurements is done according to the Bayesian framework. Here, two steps are performed alternately, namely the prediction step and the filtering step. In the prediction step, the density $f_k^e(\vx_k) := f_k^x(\vx_k|\vu_{0:k},\vz_{0:k})$ of the previous filtering step is propagated to the next time step according to
\begin{multline}
\label{eq:prediction}
f_{k+1}^p(\vx_{k+1})
:= f_{k+1}^x(\vx_{k+1}|\vu_{0:k}, \vz_{0:k}) \\
= \hspace{-1mm}\int\hspace{-1mm} \underbrace{f(\vx_{k+1}| \vx_k, \vu_k, \vw_k)}_{\delta(\vx_{k+1} - \va_k(\vx_k, \vu_k, \vw_k))}\cdot f_k^e(\vx_k) \cdot f_k^w(\vw_k) \dd \vx_k \dd\vw_k~,\hspace{-1mm}
\end{multline}
where $\vz_{0:k} = (\vz_0,\vz_1,\ldots,\vz_k)$ denotes the measurements up to and including time step $k$, $f(\vx_{k+1}| \vx_k, \vu_k, \vw_k)$ 
%
%
is the transition density depending on the dynamics model \eqref{eq:sysmodel}, and $\delta(\cdot)$ is the \DD.

The filtering step determines the \emph{posterior density} $f_k^e(\vx_k)$ of the system state $\vx_k$ based on all acquired measurement values according to \emph{Bayes' law}
\begin{equation}
\label{eq:filtering}
f_k^e(\vx_k) = c_k\cdot f(\vz_k|\vx_k)\cdot f_k^p(\vx_k)~,
\end{equation}
where $c_k$ is a normalization constant and $f(\vz_k|\vx_k)$ is the likelihood function given by
\begin{equation}
\label{eq:likelihood}
f(\vz_k|\vx_k) = \int \delta(\vz_k - \vh_k(\vx_k, \vv_k)) \cdot f_k^v(\vv_k) \dd \vv_k
\end{equation}
and the measurement model \eqref{eq:msrmodel}.

In general, for arbitrary nonlinear systems with arbitrarily distributed random vectors, there exist no analytical solutions of the prediction step and filtering step. Thus, for efficient estimation, it is inevitable to apply an approximate solution. In the following, an adaptive approximation scheme is proposed, where the predicted and posterior state densities are represented by means of Gaussian mixtures
\begin{equation}
\label{eq:gm}
f_k^\bullet(\vx_k) = \sum_{i=1}^{L_k^\bullet} \omega_{k,i}^\bullet\cdot\Gauss(\vx_k; \hvx_{k,i}^\bullet, \C_{k,i}^\bullet)~,~\bullet \in \{e,p\}
\end{equation}
where the number $L_k^\bullet$ of mixture components is variable and adapted on-line by the proposed Gaussian mixture filter.

\section{Statistical Linearization}
\label{sec:linearization}
Substituting the Gaussian mixtures representing the noise and the state density into the prediction step and the filtering step, it can be easily seen that estimation can be performed component-wise. For example in case of the prediction, using \eqref{eq:gm_noise} and \eqref{eq:gm} with $\bullet=e$ in \eqref{eq:prediction} gives rise to
\begin{multline}
\label{eq:prediction_componenetwise}
f_{k+1}^p(\vx_{k+1}) =  \sum_{i=1}^{L_k^e}\sum_{j=1}^{L_k^w} \omega_{k,i}^e\cdot \omega_{k,j}^w \cdot \biggl(\int\hspace{-1mm} f(\vx_{k+1}|\vx_k, \vu_k, \vw_k)\cdot \\
\Gauss(\vx_k; \hvx_{k,i}^e, \C_{k,i}^e) \cdot \Gauss(\vw_k; \hvw_{k,j}, \C_{k,j}^w) \dd \vx_k \dd\vw_k\biggr)~.
\end{multline}
Thus, it is sufficient to focus in the following on the simplified nonlinear transformation
\begin{equation}
\label{eq:nonlinear transformation}
\rvy = \vg(\rvx)~,
\end{equation}
which maps the Gaussian random vector $\rvx$ with density $\Gauss(\vx;\hvx, \C^x)$ to the random vector $\rvy$. This nonlinear transformation can be replaced by $\va_k(\cdot)$ in the prediction step and by $\vh_k(\cdot)$ in the filtering step, while the Gaussian random vector $\rvx$ in \eqref{eq:nonlinear transformation} represents the joint Gaussian of the state and noise.

\subsection{Classical Linearization}
\label{sec:linearization_taylor}
Calculating the density or the statistics of $\rvy$ cannot be carried out in closed form. Hence, \emph{directly} processing the density or the moments  is computationally demanding and imprecise, or even impossible. An exception are linear transformations, where the Kalman filter \cite{Kalman1960} provides analytic expressions of the Bayesian estimation problem. To apply the Kalman filter equations to nonlinear transformations, a typical way is to linearize the nonlinear transformation, which results in the extended Kalman filter \cite{Simon2006}. Here, it is assumed that the nonlinear transformation can be approximated by a linear transformation through a first-order Taylor series expansion around the mean $\hvx$. In case of mild nonlinearities the linearization error of this approximation is acceptable. However, for this type of linearization the spread of $\rvx$, i.e., the covariance matrix $\C^x$ is not taken into account and there is no measure which allows to quantify the linearization error.

\subsection{Statistical Linear Regression}
\label{sec:linearization_regression}
To overcome these flaws, deterministic sampling techniques are employed instead, which allow for propagating the mean and the covariance of $\rvx$ through the nonlinear transformation \eqref{eq:nonlinear transformation}. In doing so, linearizing the transformation by so-called statistical linear regression or \emph{statistical linearization} is possible \cite{Lefebvre2005, Vercauteren_TSP2005}. More precisely, statistical linearization calculates a matrix $\G$ and a vector $\vb$ such that
\begin{equation}
\label{eq:statistical_linearization}
\rvy = \vg(\rvx) \approx \G\cdot \rvx + \vb~,
\end{equation}
where the error term
\begin{equation}
\label{eq:error}
\rve = \vg(\rvx) - \G\cdot\rvx + \vb
\end{equation}
describes the deviation of the nonlinear transformation and its linear approximation. To determine $\G$ and $\vb$, the nonlinear transformation $\vg(\cdot)$ is evaluated at a set of weighted \emph{regression points} $\{\alpha_i, \vx_i\}_{i=1\ldots L}$ with non-negative weights $\alpha_i$ with $\textstyle \sum_i \alpha_i = 1$, which results in points $\vy_i = \vg(\vx_i)$ for $i=1\ldots L$. This set of points is chosen in such a way that the mean $\hvx$ and covariance $\C^x$ of $\rvx$ are captured exactly, that is
\begin{equation}
\label{eq:momentx}
\hvx = \sum_{i=1}^L \alpha_i\cdot \vx_i~\text{ and }~
\C^x = \sum_{i=1}^L \alpha_i\cdot (\vx_i-\hvx)\cdot(\vx_i-\hvx)\T~.
\end{equation}
Then $\G$ and $\vb$ are determined by minimizing the weighted sum of squared errors
\begin{equation}
\label{eq:calculateAb}
\{\G, \vb\} = \arg \min_{\G,\vb} \(\sum_{i=1}^L \alpha_i\cdot\ve_i\T\cdot\ve_i\)
\end{equation}
with $\ve_i = \vy_i - (\G\cdot \vx_i + \vb)$. The solution of \eqref{eq:calculateAb} is given by
\begin{equation}
\label{eq:solutionAb}
\G = \(\C^{xy}\)\T\(\C^x\)^{-1}~\text{ and }~\vb = \hvy - \G\cdot \hvx~,
\end{equation}
where the set of propagated points $\{\alpha_i, \vy_i\}_{i=1\ldots L}$ is used to approximate the mean, covariance, and cross-covariance of $\rvy$ according to
\begin{align}
\label{eq:momenty}
\hvy \approx \sum_{i=1}^L  \alpha_i\cdot \vy_i&~,\quad
\C^y \approx \sum_{i=1}^L \alpha_i\cdot (\vy_i-\hvy)\cdot(\vy_i-\hvy)\T~,\\
\C^{xy} &\approx \sum_{i=1}^L \alpha_i\cdot (\vx_i-\hvx)\cdot(\vy_i-\hvy)\T~.
\end{align}
The linearization error is characterized by the error term \eqref{eq:error} and has zero-mean and the covariance matrix
\begin{equation}
\label{eq:errorCov}
\C^e = \C^y - \G\C^x\G\T~.
\end{equation}
Thus, by means of the covariance $\C^e$ it is possible to quantify the linearization error. If $\C^e$ is a zero matrix, the density of the error $\rve$ corresponds to a \DD \cite{Parthasarathy2005} and the transformation $\vg(\cdot)$ is affine with $\vg(\rvx) = \G\cdot\rvx+\vb$.

\subsection{Calculating the Regression Points}
\label{sec:linearization_poins}
Many approaches for calculating the set of regression points have been proposed in the recent years. They differ in the number of regression points $L$ and the way these points are chosen. In the following example, both selection schemes used in this paper are briefly introduced.

\begin{example}
\label{ex:regression}
\small
In the simulations described in \Sec{sec:sim} the famous \emph{unscented transform} \cite{Julier_IEEE2004} and the \emph{Gaussian estimator} \cite{IFAC08_Huber} are considered. For both, the calculation of the sigma points $\vx_i\in\NewR^{n_x}$ can be summarized as
\begin{align}
\label{eq:regression_points}
\vx_1 &= \hvx~,\\
\vx_i &= \hvx + \nu_j\cdot\P_l\quad,\quad i=l+1+(j-1)\cdot n_x~,
\end{align}
where $\P_l$, $l=1\ldots n_x$ is the $l$th column of the matrix $\P = \sqrt{\C^x}$ and $\nu_j$,  $j=1\ldots N$ are scaling factors. This results in a number of $L = n_x\cdot N +1$ regression points. The type of the matrix root, the scaling factors, and the weights $\alpha_i$ of the regression points depend on the considered selection scheme. 

In case of the unscented transform, the Cholesky decomposition is chosen as matrix root.
For the scaling factors holds $N=2$ and $\nu_1 = \sqrt{n_x+\kappa} = -\nu_2$, where $\kappa$ is a  scaling parameter. The weights are $\alpha_1 = \tfrac{\kappa}{n_x+\kappa}$ and $\alpha_i = \tfrac{1}{2(n_x+\kappa)}$, for $i>1$. 
 
The Gaussian estimator utilizes the eigenvalue decomposition for calculating the matrix root. The number of scaling factors $N$ and thus the total number $L$ of regression points can be varied. Since the scaling factors result from solving a optimization problem, there is no closed-form expression. For $N=2$ and $N=4$, the scaling factors can be calculated to
\begin{align}
	\label{eq:nu2}
	N=2:&& &\nu_j \in\{-1.2245, 1.2245\}~, \\
	\label{eq:nu4}
	N=4:&& &\nu_j \in\{-1.4795, -0.5578, 0.5578, 1.4795\}~.\quad
\end{align}
%
%
The regression points are equally weighted with $\forall i: \alpha_i = \tfrac{1}{L}$\,.
\end{example}

It is important to note that the proposed adaptive Gaussian mixture filter is not restricted to these two selection schemes.  In fact, any selection scheme for statistical linearization including those described in \cite{Arasaratnam_TAC2009, Ito_TAC2000, Schei_Automatica1997} can be used, depending on the considered application as well as the desired estimation performance and computational demand.

\section{Splitting Scheme}
\label{sec:splitting}
Given a random vector $\rvx$, whose density function $f^x(\vx)$ is a Gaussian mixture with $L^x$ components according to \eqref{eq:gm}, it is possible to linearize the nonlinear transformation $\vg(\cdot)$ for each component of $f^x(\vx)$. This kind of component-wise or \emph{local linearization} leads to an improved approximation of the true density function $f^y(\vy)$ of $\rvy$  compared to a single, global linearization. To further improve the approximation, especially in case of strong nonlinearities and/or large variances of some components, the idea is to select a component of $f^x(\vx)$ and split it into several components with reduced weights and covariances. It was demonstrated for example in \cite{Alspach_GaussianSumApproximation} that the filtering accuracy of local linearization approaches benefits from this decrease of the covariances and simultaneous increase of the number of Gaussians.

\subsection{Component Selection}
\label{sec:splitting_component}
A straightforward way to select a Gaussian component for splitting is to consider the weights $\omega_i^x$, $i=1\ldots L^x$. The component with the highest weight is then split. This however does not take the nonlinearity of $\vg(\cdot)$ in the support 
of the selected component into account. Since linearization is performed component-wise and locally, a more reasonable selection would be to consider also the induced linearization error of each component. For this purpose, statistical linearization already provides an appropriate measure for the linearization error in form of the covariance matrix $\C^e$ in \eqref{eq:errorCov}.

In order to easily assess the linearization error in the multi-dimensional case, the trace operator is applied to $\C^e$, which gives the measure
\begin{equation}
\label{eq:trace}
	\epsilon = \trace\(\C^e\) \in [0, \infty)~.
\end{equation}
Geometrically speaking, the trace is proportional to circumference of the covariance ellipsoid  corresponding to $\C^e$. The larger $\C^e$ and thus the linearization error, the larger is $\epsilon$. Conversely, the trace is zero, if and only if $\C^e$ is the zero matrix, i.e., $\epsilon = 0 \Leftrightarrow \C^e = \mat 0$. Hence, \eqref{eq:trace} is only zero, when there is no linearization error, that is, the nonlinear transformation $\vg(\cdot)$ is affine in the support of the considered Gaussian component.

Besides the linearization error, the contribution of a component to the nonlinear transformation is important as well.  That is, the probability mass of the component, which is given by its weight $\omega^x_i$, has also to be taken into account. This avoids splitting irrelevant components. Putting all together the criterion 
for selecting a component $i$ for splitting is defined as
\begin{equation}
\label{eq:splitting_criterion}
	s_i = \(\omega_i^x\)^\gamma \cdot \(1-\exp\(-\epsilon_i\)\)^{1-\gamma} \in [0,1]
\end{equation}
for $i=1\ldots L^x$, where $1-\exp\(-\epsilon_i\)$ normalizes the linearization measure \eqref{eq:trace} into the interval $[0,1]$. For a geometric interpolation between weight and linearization error of component $i$, the parameter $\gamma \in [0, 1]$ used. With $\gamma=0$, selecting a component for splitting only focuses on the linearization error, while $\gamma=1$ considers the weight only.

Component selection criteria for splitting have also been proposed in \cite{Faubel_ICASSP2010, Rauh_CCA09}. The criterion in \cite{Faubel_ICASSP2010} is designed for the unscented transform only, while the criterion in \cite{Rauh_CCA09} can only be calculated analytically in some special cases. The proposed criterion instead is generally applicable.

\subsection{Splitting a Gaussian}
\label{sec:splitting_gaussian}
Assume that according to the selection criterion \eqref{eq:splitting_criterion}, the Gaussian component $\omega\cdot\Gauss(\vx; \hvx, \C^x)$ is chosen. Splitting this Gaussian into many can be formulated as replacing the Gaussian by a Gaussian mixture according to 
\begin{equation}
\label{eq:gmsplit}
\omega\cdot\Gauss(\vx; \hvx, \C^x) \approx \sum_{j=1}^L \omega_j \cdot \Gauss(\vx;\hvx_j, \C_j)~.
\end{equation}
It can be easily verified that for $L>1$, the number of free parameters, i.e., weights, mean vectors and covariance matrices, is larger than the number of given parameters. More precisely, splitting a Gaussian is an ill-posed problem. In order to reduce the degrees of freedom and to not introduce errors concerning the mean and covariance, splitting is performed in a moment-preserving fashion. Thus, it must hold that 
\begin{align}
	\begin{split}
	\label{eq:momentpreserving}
	\omega &= \sum_{j=1}^L \omega_j~\quad ,\quad~
	\hvx = \sum_{j=1}^L \tfrac{\omega_j}{\omega}\cdot\hvx_j~,\\
	\C^x &= \sum_{j=1}^L \tfrac{\omega_j}{\omega} \cdot \(\C_j + \hvx_j\,\hvx_j\T\) - \hvx\, \hvx\T~.
	\end{split}
\end{align}
%
To further simplify the problem, splitting is restricted in direction of the eigenvectors of $\C^x$, which is computationally cheap and numerically stable. Furthermore, it reduces the problem to splitting a univariate standard Gaussian.

\subsubsection{Univariate standard Gaussian}
\label{sec:splitting_gaussian_univariate}
A moment-preserving split of a univariate standard Gaussian $\Gauss(x; 0, 1)$ into a mixture with $L$ components requires to determine $3\cdot L$ free parameters. By forcing symmetry, i.e., the means $\hx_j$ are placed symmetrically around the mean $\hx$ with symmetrically chosen weights and for the variances holds $\forall j: \sigma_j^2=\sigma$, the number of free parameters is reduced to $L+1$. In \cite{MFI08_HuberBailey}, a splitting library with symmetric components is proposed. Unfortunately, preserving the moments is not guaranteed. Instead, the following split into two components is used throughout this paper.

\begin{example}
\small
Following the approach proposed in \cite{Faubel_SPL2009}, the univariate standard Gaussian is split into the mixture $\tfrac{1}{2}\cdot \Gauss(x; \hx, \sigma^2) + \tfrac{1}{2}\cdot \Gauss(x; -\hx, \sigma^2)$. 
The moment-preserving constraints of splitting \eqref{eq:momentpreserving} 
lead to the dependency
%
$\sigma^2 = 1-\hx^2$ 
%
between $\hx$ and $\sigma$, where $\hx$ is now the only free parameter. This equation is valid for $\hx \in [-1,1]$ and contains the trivial solution $\hx=0$. 
Generally, $\hx$ may be determined dynamically by minimizing the resulting linearization error. But throughout this paper, $\hx$ is set to $0.5$ for simplicity.
\end{example}
\vspace{-.5mm}

To determine the parameters of more than two components, additional constraints, e.g., capturing higher order moments like the skewness or the kurtosis have to be considered additionally. Since splitting is performed recursively in this paper (see \Sec{sec:gaussianmixture}), the new introduced components can be split in the subsequent splitting step if the local linearization error may not be reduced sufficiently. Splitting into two components is a good compromise between reducing the linearization error on the one hand and controlling the growth of the number of  components and the computational load on the other~hand. 

\subsubsection{Multivariate Gaussian}
\label{sec:splitting_gaussian_multivariate}
Applying univariate splitting to the multivariate case requires the eigenvalue decomposition of the covariance matrix $\C^x = \V\D\V\T$, with $\V$ being the matrix of eigenvectors and $\D$ being the diagonal matrix of eigenvalues according to
\begin{equation}
\label{eq:eigenvalue}
\V = 
\begin{bmatrix}
\vv_1 & \vv_2 & \hdots & \vv_{n_x}
\end{bmatrix}~,~
\D = \diag\(\lambda_1, \lambda_2, \ldots, \lambda_{n_x}\)~,
\vspace{-2mm}
\end{equation}
where $\vv_i\in\NewR^{n_x}$ are the (orthonormal) eigenvectors and $\lambda_i$ are the eigenvalues. 
$\V$ is a rotation matrix and the eigenvalue decomposition of $\C^x$ corresponds to the transformation 
%
\begin{equation}
\label{eq:eigentransformation}
\rvx = \V\cdot \rvz
\vspace{-.5mm}
\end{equation}
of a Gaussian random vector $\rvz$ with density 
\begin{equation}
	\label{eq:diagz}
	f^z(\vz) = \Gauss(\vz; \hvz, \D) 
	= \prod_{i=1}^{n_x} \Gauss(z_i; \hz_i, \lambda_i)
\end{equation}
to a Gaussian random vector $\rvx$ with density $\Gauss(\vx;\hvx,\C^x)$\,. 
Since the Gaussian $f^z(\vz)$ has a diagonal covariance matrix, the eigenvectors are parallel to the axes of the coordinate system. Thus, univariate splitting can be easily applied along the eigenvectors by replacing a univariate Gaussian on the right-hand side of \eqref{eq:diagz} by a Gaussian mixture. 

Assuming that eigenvector $\vv_l$ is chosen for splitting and let $\textstyle \sum_{j=1}^L \omega_j'\cdot \Gauss(z_l; \hz_j', \sigma_j^2)$ be the Gaussian mixture that approximates a univariate standard Gaussian as described above. As this mixture approximates a standard Gaussian, its components have to be shifted by adding $\hz_l$ and scaled by multiplying with $\sqrt{\lambda_l}$ in order to match the mean $\hz_l$ and the variance $\lambda_l$, respectively. These operations result in \vspace{-1mm}
\begin{equation}
\label{eq:scaling}
\Gauss(z_l; \hz_l, \lambda_l)\approx\sum_{j=1}^L \omega_j'\cdot \Gauss\(z_l; \hz_l + \sqrt{\lambda_l}\hz_j', \lambda_l\sigma_j^2\)~. 
\end{equation}
Plugging \eqref{eq:scaling} into \eqref{eq:diagz} leads to\vspace{-1mm}
\begin{multline}
\label{eq:univariatemixture}
\Gauss(\vz; \hvz, \D) 
	\approx \\ \sum_{j=1}^L \omega_j'\cdot \Gauss\(z_l; \hz_l + \sqrt{\lambda_l}\hz_j', \lambda_l\sigma_j^2\) \cdot \prod_{\substack{i=1 \\ i\neq l}}^{n_x} \Gauss\(z_i; \hz_i, \lambda_i\) ~.
\end{multline}
Transforming this mixture via \eqref{eq:eigentransformation} gives the desired splitting result \eqref{eq:gmsplit} with the weights, means, and covariance matrices
\begin{align}
\begin{split}
\label{eq:spitmoments}
\omega_j &= \omega\cdot \omega_j'~,\\
\hvx_j &= \hvx + \sqrt{\lambda_l}\cdot\hz_j'\cdot\vv_l~,\\
\C_j &= \C^x + \lambda_l\cdot(\sigma_j^2-1)\cdot\vv_l\vv_l\T~,
\end{split}
\end{align}
for $j=1\ldots L$. 

It is worth mentioning that the calculation of the parameters in \eqref{eq:spitmoments} is independent of the number of components $L$ and does not necessarily require a symmetric, moment-preserving splitting. Thus, arbitrary splitting methods of univariate standard Gaussians besides those described in this paper, can be used with these formulae.

\subsection{Splitting Direction}
\label{sec:splitting_direction}
So far, no criterion for selecting an appropriate eigenvector for splitting is defined. A straightforward criterion may be the eigenvector with the largest eigenvalue as in \cite{Faubel_SPL2009, MFI08_HuberBailey}. 
But since \eqref{eq:splitting_criterion} determines the Gaussian component that causes the largest linearization error, merely splitting along the eigenvector with the largest eigenvalue does not  take this error into account. 

The key idea of the proposed criterion is to evaluate the deviation between the nonlinear transformation \eqref{eq:nonlinear transformation} and its linearized version \eqref{eq:statistical_linearization} along each eigenvector. The eigenvector with the largest deviation is then considered for splitting, i.e., the Gaussian is split in direction of the largest deviation in order to cover this direction with more Gaussians, which will reduce the error in subsequent linearization steps.

By means of the error term \eqref{eq:error}, the desired criterion for the splitting direction is defined as
\begin{align}
\label{eq:deviationalongeigenvector}
d_l &:= \int_\NewR \rve(\vx_l(\nu))\T\cdot\rve(\vx_l(\nu))\cdot \Gauss(\vx_l(\nu);\hvx, \C^x) \dd \nu
\end{align}
with $\vx_l(\nu) := \hvx+\nu\cdot \vv_l$, $l=1\ldots n_x$, and $\vv_l$ being the $l$th eigenvector $\C^x$. The integral in \eqref{eq:deviationalongeigenvector} cumulates the squared deviations along the $l$th eigenvector under the consideration of the probability at each point $\vx_l(\nu)$. The eigenvector that maximizes \eqref{eq:deviationalongeigenvector} is then chosen for splitting. 
Unfortunately, due to the nonlinear transformation $\vg(\cdot)$ in \eqref{eq:error}, this integral cannot be solved in closed-form in general. For an efficient and approximate solution, 
the regression point calculation schemes described in \Sec{sec:linearization_poins} are employed to approximate the Gaussian in \eqref{eq:deviationalongeigenvector} in direction of $\vv_l$ by means of a Dirac mixture. This automatically leads to a discretization of the integral at a few but carefully chosen points.


\begin{figure}[tb]%
\begin{tikzpicture}
\node[rblock, text width=1.7cm] at (-.1,0) (linp) {\small Linearization};
\node[dblock, text width=.7cm, inner ysep=0mm, fill=lightgray] at (2,0) (stopp) {stop?};
\node[rblock] at (2,1.5) (splitp) {\small Splitting};
\node[rblock] at (4,0) (pred) {\small Prediction};
\node[rblock] at (6,0) (redp) {\small Reduction};

\node[rblock] at (-.3,-3) (rede) {\small Reduction};
\node[rblock] at (1.7,-3) (filt) {\small Filtering};
\node[rblock] at (3.7,-1.5) (splite) {\small Splitting};
\node[dblock, text width=.7cm, inner ysep=0mm, fill=lightgray] at (3.7,-3) (stope) {stop?};
\node[rblock, text width=1.7cm] at (5.8,-3) (line) {\small Linearization};

\node[rblock, text width=1.2cm, inner ysep=1mm] at (6,-1.55) (delay) {\footnotesize $k+1\rightarrow k$};

\draw[ablock]	(linp) -- (stopp);
\draw[ablock]	(stopp) -- (splitp) node[right, near start] () {\ding{53}};
\draw[ablock]	(stopp) -- (pred) node[above, near start] () {\ding{51}};
\draw[ablock]	(pred) -- (redp);
\draw[ablock, latex-]	(linp.128) |- (splitp);
\draw[ablock]	(redp) -- (redp |- delay.north) node[right,pos=.57] () {$f_{k+1}^p$};

\draw[ablock]	(line) -- (stope);
\draw[ablock]	(filt) -- (rede);
\draw[ablock]	(stope) -- (filt) node[above, at start] () {\ding{51}};
\draw[ablock]	(stope) -- (splite) node[right, near start] () {\ding{53}};
\draw[ablock]	(splite) -| (line.160);
\draw[ablock]	(rede) -- (rede |- linp.south) node[right,pos=.75] () {$f_{k}^e$};
\draw[ablock]	(delay) -- (delay |- line.north) node[right,pos=.4] () {$f_{k}^p$};
\draw[ablock,latex-]	(linp.west) -- +(-.5,0)  node[above,near end] () {$f_{0}^x$};
\draw[ablock,latex-] (filt.north) -- +(0,.4) node[above,pos=.8] () {$\vz_k$};

\node at (5,1.5) () {\emph{Prediction step}};
\node at (1.5,-1.5) () {\emph{Filtering step}};

\node[fitblock, fill=none, dashed, fit=(linp) (splitp) (redp) (stopp)] (prediction) {};
\node[fitblock, fill=none, dashed, fit=(line) (splite) (rede) (stope)] (prediction) {};
\end{tikzpicture}

\caption{Flow chart of the proposed adaptive Gaussian mixture filter. Both the prediction and filtering step employ splitting and reduction for adapting the number of mixture components.}%
\label{fig:agmf}%
\end{figure}
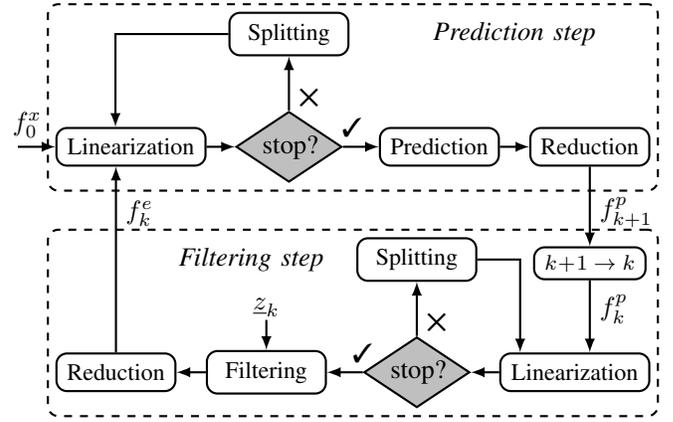

\section{Adaptive Gaussian Mixture Filter}
\label{sec:gaussianmixture}
Based on the statistical linearization described in \Sec{sec:linearization} and the splitting procedure proposed in \Sec{sec:splitting}, the complete adaptive Gaussian mixture filter (AGMF) is now derived. 
The key idea of AGMF is to dynamically increase the number of Gaussians of a given mixture at regions with large linearization errors. The number is reduced after each prediction and filtering in order to limit the computational and memory demand.

\subsection{Prediction Step}
\label{sec:gaussianmixture_prediction}
The major operations to be performed in the prediction step are illustrated in the upper part of \Fig{fig:agmf}. The following paragraphs provide detailed descriptions of these operations.

\subsubsection{Linearization}
\label{sec:gaussianmixture_prediction_linearization}
As shown in \eqref{eq:prediction_componenetwise} the prediction step can be performed component-wise. Therefore, the nonlinear system function is linearized statistically at the weighted joint Gaussian $\omega_{k,s} \cdot \Gauss(\vX_k; \hvX_{k,s}, \C^X_{k,s})$ with $\rvX_k = [\rvx_k\T, \rvw_k\T]\T$, $s=(i-1)\cdot L_k^w+j =1\ldots L_k^e\cdot L_k^w$, $\omega_{k,s} = \omega_{k,i}^e\cdot \omega_{k,j}^w$, and 
\begin{equation}
\label{eq:}
\hvX_{k,s} = \begin{bmatrix}
\hvx_{k,i}^e \\ \hvw_{k,j}
\end{bmatrix}~,\quad
\C^X_{k,s} = 
\begin{bmatrix}
\C^e_{k,i} & \mat 0 \\
\mat 0 & \C^w_{k,j}
\end{bmatrix}~.
\end{equation}
With \eqref{eq:solutionAb}, the linearization results in
\begin{equation}
\label{eq:linearizedsystem}
\rvx_{k+1} = \underbrace{\begin{bmatrix}\A_{k,s} & \B_{k,s}\end{bmatrix}}_{=\, \mat G_{k,s}}\cdot \rvX_k + \vb_{k,s}~.
\end{equation}

\subsubsection{Splitting}
\label{sec:gaussianmixture_splitting}
Due to the nonlinearity of the system function $\va_k(\cdot)$, some of the mixture components $\omega_{k,s} \cdot \Gauss(\vX_k; \hvX_{k,s}, \C^X_{k,s})$ may locally cause severe linearization errors. 
These errors are quantified by means of the selection criterion \eqref{eq:splitting_criterion}. The component maximizing \eqref{eq:splitting_criterion} will be split in direction of the largest deviation between the nonlinear function $\va_k(\cdot)$ and its linearized version \eqref{eq:linearizedsystem} as described in \Sec{sec:splitting_direction}. 
After splitting this Gaussian, linearization is performed for the newly introduced mixture components. The linearization need not to be repeated for the remaining mixture components as they are not affected by the splitting. 

Splitting Gaussians and the subsequent linearization is repeated until a stopping criterion is satisfied. This stopping criterion combines three user-defined thresholds: 
\begin{enumerate}
	\item For \emph{each} component the value of the selection criterion \eqref{eq:splitting_criterion} shall drop below the \emph{error threshold} $\epsilon_\mathrm{max} \in [0,1]$.
	\item The number of Gaussians shall not grow beyond the \emph{component threshold} $L_\mathrm{max}$. 
	\item The deviation between the original Gaussian mixture $f(\vx)$ and the mixture obtained via splitting $\tilde f(\vx)$ shall remain below a \emph{deviation threshold} $d_\mathrm{max} \in[0,1]$.
\end{enumerate}

In the latter case, the deviation is determined by means of the normalized integral squared distance measure~\cite{Fusion08_Huber_PGMR}
\begin{equation}
\label{eq:isd}
D\bigl(f(\vx), \tilde f(\vx)\bigr) = \frac{\int \bigl(f(\vx)-\tilde f(\vx)\bigr)^2 \dd \vx}{\int f(\vx)^2\dd \vx + \int \tilde f(\vx)^2\dd \vx} \in[0,1]~.
\end{equation}
Since splitting always introduces an approximation error to the original mixture $f(\vx)$, tracking the deviation during splitting and keeping the deviation below the threshold $d_\mathrm{max}$ avoids that errors introduced by splitting neutralize the gain in linearization. 
Splitting stops, if at least one threshold is reached.

\subsubsection{Prediction}
\label{sec:gaussianmixture_prediction_prediction}
Let $\omega_{k,\tilde s} \cdot \Gauss(\vX_k; \hvX_{k,\tilde s}, \C^X_{k,\tilde s})$ be the Gaussians resulting from the splitting step, with $\tilde s = 1\ldots \tilde L_k^p$ and $\tilde L_k^p \gg L_k^e\cdot L_k^w$. Based on these Gaussians and their corresponding locally linearized system models \eqref{eq:linearizedsystem}, the parameters of each component of the predicted Gaussian mixture $f_{k+1}^p(\vx_{k+1})$ can be calculated by means of the Kalman predictor according to
\begin{align}
\begin{split}
\label{eq:prediction_parameters}
\omega_{k+1,\tilde s}^p & = \omega_{k,\tilde s}~,\\
\hvx_{k+1,\tilde s}^p &= \A_{k,\tilde s} \cdot \hvx_{k,\tilde s}^e + \B_{k,\tilde s}\cdot \hvw_{k,\tilde s} + \vb_{k,\tilde s}~,\\
\C^p_{k+1,\tilde s} &= \A_{k,\tilde s}\C^e_{k,\tilde s}\A_{k,\tilde s}\T + \B_{k,\tilde s}\C_{k,\tilde s}^w\B_{k,\tilde s}\T + \C_{k,\tilde s}~,
\end{split}
\end{align}
where $\C_{k,\tilde s}$ is the linearization error covariance \eqref{eq:errorCov}.

\subsubsection{Reduction}
\label{sec:gaussianmixture_reduction}
The number of components $\tilde L_k^p$ in $f_{k+1}^p(\vx_{k+1})$ grows due to the multiplication of the Gaussian mixtures $f_k^e(\vx_k)$ and $f_k^w(\vw_k)$ for prediction and due to splitting. 
It is necessary to bound this growth in order to reduce the computational and memory demand of subsequent prediction and filtering steps. For this purpose, one can exploit the redundancy and similarity of Gaussian components. Furthermore, many components will have weights close to zero, thus they can be removed without introducing significant errors. 
To reduce a Gaussian mixture, many algorithms have been proposed in the recent years (see for example \cite{Fusion08_Huber_PGMR, Runnalls2007, Salmond_SPIE_1990, West1993_JSTOR}). 
Most of these algorithms require a \emph{reduction threshold} $L_{k+1}^p$--typically much smaller than $L_\mathrm{max}$--to which the number of components of the given Gaussian mixture has to be reduced. 
In the simulations, Runnalls' reduction algorithm \cite{Runnalls2007} is employed as it provides a good trade-off between computational demand and reduction errors.

With the reduction to $L_{k+1}^p$ components, the calculation of the predicted Gaussian mixture $f_{k+1}^p(\vx_{k+1})$ in \eqref{eq:gm} is finished.

\subsection{Filtering Step}
\label{sec:gaussianmixture_filtering}
The operations to be performed for the filtering step are almost identical to the prediction step (see \Fig{fig:agmf}). Thus, only linearization and filtering are described in the following. Splitting and reduction coincide with the prediction step.

\subsubsection{Linearization}
Linearization and filtering are also performed component-wise. Let $\omega_{k,s}\cdot \Gauss(\vX_k, \hvX_{k,s}, \C_{k,s}^X)$ be the joint Gaussian comprising the $i$th component of the predicted mixture $f_k^p(\vx_k)$ and the $j$th component of the measurement noise mixture \eqref{eq:gm_msrnoise}, where $s=(i-1)\cdot L_k^v+j=1\ldots L_k^p\cdot L_k^v$\,. 
The corresponding linearized measurement model is
\begin{equation}
\label{eq:linearizedmsr}
\rvz_k = \begin{bmatrix}\H_{k,s} & \D_{k,s}\end{bmatrix}\cdot \rvX_k + \vb_{k,s}
\end{equation}
with joint state $\rvX_k = \begin{bmatrix}\rvx_k\T, \rvv_k\T\end{bmatrix}\T$.

\subsubsection{Filtering}
Let $\omega_{k,\tilde s} \cdot \Gauss(\vX_k; \hvX_{k,\tilde s}, \C^X_{k,\tilde s})$ be the Gaussians resulting from splitting, with $\tilde s = 1\ldots \tilde L_k^e$ and $\tilde L_k^e\gg L_k^p\cdot L_k^v$.
Given the current measurement value $\vz_k$, the Kalman filter update equations applied on these Gaussians and their corresponding locally linearized measurement models \eqref{eq:linearizedmsr} give rise to the parameters of each component of the posterior Gaussian mixture $f_{k}^e(\vx_{k})$ 
\begin{align}
\begin{split}
\label{eq:fitlering_parameters}
\omega_{k,\tilde s}^e & = c_k\cdot \omega_{k,\tilde s}\cdot\Gauss(\vz_k;\hvz_{k,\tilde s},\S_{k,\tilde s})~,\\
\hvx_{k,\tilde s}^e &= \hvx_{k,\tilde s}^p + \K_{k,\tilde s}\(\vz_k - \hvz_{k,\tilde s}\)~,\\
\C^e_{k,\tilde s} &= \C^p_{k,\tilde s} - \K_{k,\tilde s}\H_{k,\tilde s}\C^p_{k,\tilde s}~,
\end{split}
\end{align}
with 
predicted measurement $\hvz_{k,\tilde s} = \H_{k,\tilde s}\cdot\hvx_{k,\tilde s}^p + \D_{k,\tilde s}\cdot\hvv_{k,\tilde s} + \vb_{k,\tilde s}$, 
Kalman gain $\K_{k,\tilde s} = \C^p_{k,\tilde s}\H_{k,\tilde s}\T\S_{k,\tilde s}^{-1}$, 
innovation covariance $\S_{k,\tilde s} = \H_{k,\tilde s}\C_{k,\tilde s}^p\H_{k,\tilde s}\T + \D_{k,\tilde s}\C_{k,\tilde s}^v\D_{k,\tilde s}\T + \C_{k,\tilde s}$, and $\C_{k,\tilde s}$ being the linearization error covariance \eqref{eq:errorCov}. The calculation of the weight $\omega_{k,\tilde s}^e$ in \eqref{eq:fitlering_parameters} is adapted from \cite{Alspach_GaussianSumApproximation, Simandl_IFAC2005}, where $\textstyle c_k = 1/\sum_{\tilde s} \omega_{k,\tilde s}\cdot\Gauss(\vz_k;\hvz_{k,\tilde s},\S_{k,\tilde s})$ is a normalization constant.

After the reduction to $L_k^e$ components, the posterior Gaussian mixture $f_{k}^e(\vx_{k})$ in \eqref{eq:gm} is completely determined.

%
%
            
\begin{table}[b]%
\caption{Approximation error (KLD $\times$ $10$) for different splitting schemes and numbers of components.}
\label{tab:sim_shape}
\centering
\begin{tabular}{|c||c|c|c|c|c|c|c|}
\hline
splitting & \multicolumn{7}{c|}{number of Gaussians}
\\
scheme & 1 & 2 & 4 & 8 & 16 & 32 & 64
\\ \hline
max. eigenvalue & 2.01 & 0.77 & 0.64 & 0.47 & 0.39 & 0.21 & 0.26
\\
$\gamma = 1$ & 2.01 & 0.77 & 0.59 & 0.34 & 0.20 & 0.12 & 0.07
\\
$\gamma = 0.5$ & 2.01 & 0.77 & 0.40 & 0.22 & 0.07 & 0.03 & 0.02
\\\hline
\end{tabular}
\end{table}

\section{Simulation Results}
\label{sec:sim}
Two numerical simulations are conducted in order to demonstrate the performance of the proposed AGMF. 

\subsection{Shape Approximation}
\label{sec:sim_shape}
In the first simulation, the nonlinear growth process 
\begin{equation}
\label{eq:sim_shape}
\y = g(\rvx) = \frac{\rv \xi}{2} + 5\cdot \frac{\rv \xi}{1 + \rv \xi^2} + \w
\end{equation}
adapted from \cite{Kitagawa1996} is considered, where $\rvx = [\rv \xi, \w]\T \sim f^x(\vx) = \Gauss(\vx; [1, 0]\T, \I_2)$ with $\I_n$ being the $n\times n$ identity matrix. To approximate the density of $\y$, the Gaussian $f^x(\vx)$ is split recursively into a Gaussian mixture, where the number of components is always doubled until a maximum of $64$ components is reached. No mixture reduction and no thresholds $\epsilon_\mathrm{max}$, $d_\mathrm{max}$ are used. The Gaussian estimator with $4$ scaling factors according to \eqref{eq:nu4} is employed for statistical linearization. The true density of $\y$ is calculated via numerical integration.

Two different values for the parameter $\gamma$ of the selection criterion \eqref{eq:splitting_criterion} are used: $\gamma=0.5$, which makes no preference between the component weight and the linearization error and $\gamma=1$, which considers the weight only. Furthermore, a rather simple selection criterion is considered for comparison, where selecting a Gaussian for splitting is based on the weights only (as it is the case for $\gamma=1$), while the splitting is performed in direction of the eigenvector with the largest eigenvalue.

\Tab{tab:sim_shape} shows the Kullback-Leibler divergence (KLD, \cite{Cover1991}) between the true density of $\y$ and the approximations obtained by splitting. The approximations of the proposed splitting scheme are significantly better than the approximations of the largest eigenvalue scheme. This follows from the fact that the proposed scheme not only considers the spread of a component. It also takes the linearization errors into account. In doing so, the Gaussians are always split along the eigenvector that is closest to $\rv \xi$, since this variable is transformed nonlinearly, while $\w$ is not. This is different for the largest eigenvalue scheme, which wastes nearly half of the splits on $\w$.

The inferior approximation quality for $\gamma=1$ compared to $\gamma=0.5$ results from splitting components, which may have a high importance due to their weight but which do not cause severe linearization errors. Thus, splitting these components will not improve the approximation quality much. 

\begin{figure}[tb]%
\includegraphics[]{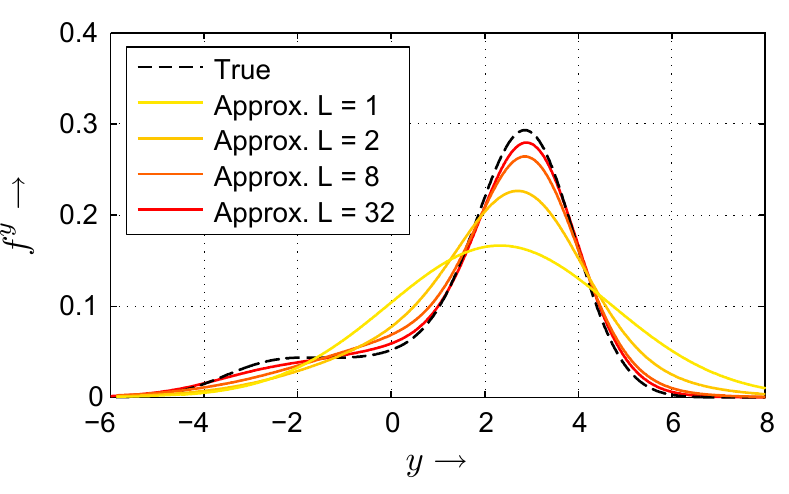}%
\caption{True density function of $\y$ (black, dashed) and approximations with an increasing number of mixture components.}%
\label{fig:sim_shape}%
\vspace{-2mm}
\end{figure}

In \Fig{fig:sim_shape}, the approximate density of $\y$ is depicted for different numbers of mixture components for $\gamma=0.5$. With an increasing number of components, the approximation \linebreak approaches the true density very well.

\begin{figure*}[tb]%
\includegraphics[]{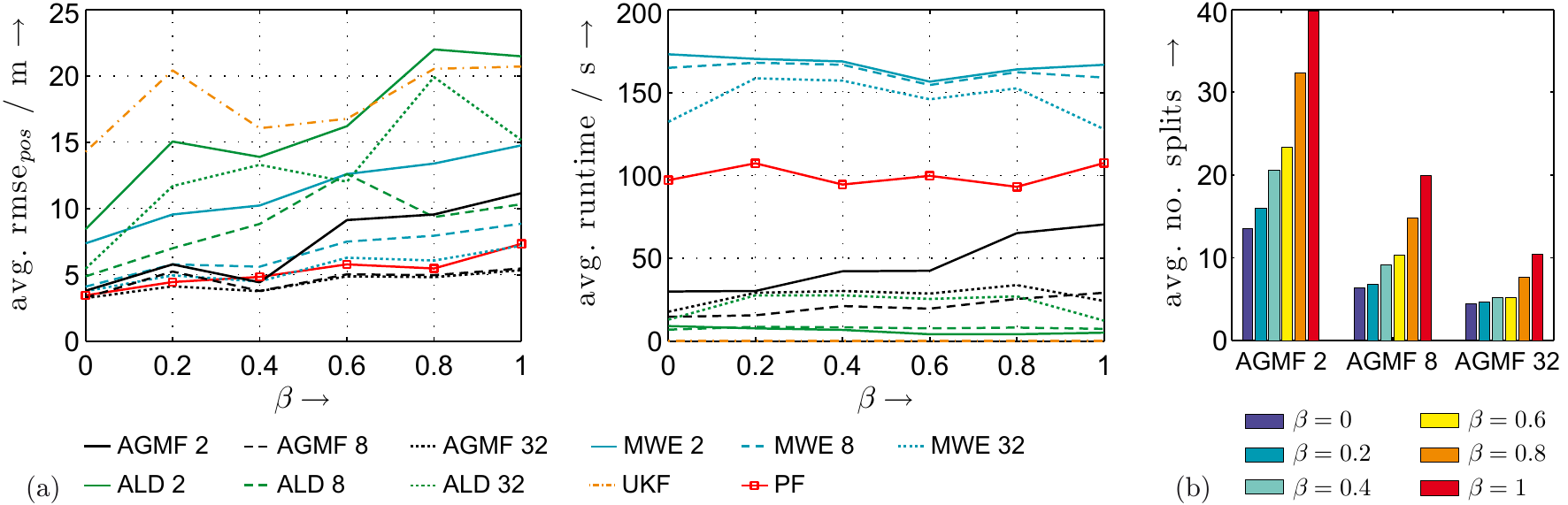}%
\caption{(a) Average rmse over all simulation runs and average runtime per simulation run for each $\beta$. (b) Average number of splits performed per prediction or filtering step of the AGMF for each $\beta$.}%
\label{fig:simND}%
\vspace{-1mm}
\end{figure*}

\subsection{Object Tracking}
\label{sec:sim_tracking}
For the second simulation example, a object tracking scenario is considered. The kinematics of the mobile object are modeled by means of the bicycle model
\begin{equation}
\label{eq:kinematics}
\rvx_{k+1} :=
\begin{bmatrix}
	\x_{k+1} \\ \y_{k+1} \\ \rv \phi_{k+1}
\end{bmatrix}
= \rvx_k + 
\begin{bmatrix}
\cos(\rv \phi_k) \\ \sin(\rv \phi_k) \\ u_k
\end{bmatrix}
+ \rvw_k~,
\end{equation}
where the system state $\rvx_k$ comprises the position $[\x_k, \y_k]\T$ and the orientation $\rv \phi_k$ of the bicycle. 
At time step $k=0$, the initial estimate of the state $\rvx_0$ is represented by a Gaussian density with mean $\hvx_0=[\Meter{100}{}, \Meter{100}{}, \Rad{0}{}]\T$ and covariance matrix $\C_0^x=\diag([10^2, 10^2, \pi^2])$. 
The system input $u_k :=\tan(\alpha_k)$ with $\alpha_k$ being the steering angle, is chosen randomly and uniformly distributed from the interval $[-0.2, 0.2]$ at each time step. The system noise $\rvw_k$ is zero-mean Gaussian with covariance matrix $\C_k^w = \diag(0.1^2, 0.1^2, 0.01^2)$. 

A radar sensor with measurement model 
\begin{equation}
\label{eq:radar}
\rvz_k =
\begin{bmatrix}
\sqrt{\x_k^2 + \y_k^2} \\ \arctan\(\y_k/\x_k\)
\end{bmatrix}
+ \rvv_k
\end{equation}
is employed for observing the object, where the measurement noise $\rvv_k$ is modeled as unimodal glint noise \cite{Wu_TAES1993} with density $f_k^v(\vv_k) = (1-\beta)\cdot\Gauss(\vv_k;\vec 0, \C_{k,1}^v) + \beta\cdot\Gauss(\vv_k;\vec 0, \C_{k,2}^v)$ with covariances $\C_{k,1}^v = \diag(1^2, 0.1^2)$ and $\C_{k,2}^v = \diag(2^2, 0.2^2)$. The parameter $\beta$ refers to the glint noise probability. Six probability values $\beta=\{0, 0.2, \ldots, 1\}$ are exploited for simulation. By increasing $\beta$ it is possible to investigate the performance of the filters for stronger noise, which is also heavily tailed for $\beta \neq 0$ and $\beta \neq 1$.

For this simulation setup, AGMF is applied with parameter $\gamma = 0.5$, error threshold $\epsilon_\mathrm{max} = 0.05$, deviation threshold $d_\mathrm{max} = 1$, and component threshold $L_\mathrm{max} = 128$ for both prediction and filtering. Three values of reduction thresholds are used, $L_k^p = L_k^e = 2, 8, 32$. For comparison, a Gaussian mixture filter (denoted as MWE) employing the simple largest-weight-largest-eigenvalue-criterion as described in the previous section is considered. Further, the adaptive level of detail (ALD) Gaussian mixture filter proposed in \cite{Faubel_ICASSP2010} is employed as well. Since ALD is only designed for the unscented transform (see Example~\ref{ex:regression}), this statistical linearization method is also used for AGMF to allow a fair comparison. The scaling parameter $\kappa$ of the unscented transform is set to $0.5$, i.e., all regression points are equally weighted. MWE and ALD use the same parameters as AGMF, except that MWE always splits until $L_\mathrm{max}$ is reached since it exploits no linearization errors.

Besides these Gaussian mixture filters, a particle filter (PF) with residual resampling and $10,000$ samples as well as the unscented Kalman filter (UKF,~\cite{Wan_2000}) with $\kappa=0.5$ are also applied.

For each glint probability and each reduction threshold, $50$ Monte Carlo simulation runs are performed, where the object is observed for $100$ time steps. In \Fig{fig:simND}\,(a), the average root means square error (rmse) of the position and the average runtime per simulation run are depicted.
The AGMFs with $8$ and $32$ components provide the best tracking performance. The PF is close to AGMF, but with a significantly higher runtime. Conversely, the UKF is by far the fastest algorithm, but leads to diverging estimates.

The splitting criterion used for ALD selects components that exhibit a high degree of nonlinearity. But splitting is performed merely in direction of the largest eigenvalue. This explains the relative poor tracking performance of ALD. 

Even if MWE is allowed to split until $L_\mathrm{max}$ is reached, the performance of MWE is always inferior to AGMF. This is due to wasting many splits, e.g., in the prediction step only one quarter and less of the splits is used for $\rv \phi_k$, which is the only nonlinearly transformed variable. 
Here, AGMF is much more effective thanks to the novel splitting criterion. Besides splitting mainly in direction of the nonlinearity, it does not require all available splits as shown in \Fig{fig:simND}\,(b). The maximum number of splits is $L_\mathrm{max} - L_k^p$ in the prediction step and analogously in the filtering step. But at most $40$ splits are performed in case of the strongest noise and when the state mixture is reduced to two components. If more components are allowed to represent the state density, the number of splits decreases as 
the approximation before splitting is already of high quality. 
This also reduces the runtime as can be seen when comparing for example AGMF~32 with AGMF~2. Here, the time consuming splitting operation has to be performed less often and the reduction operation has to reduce a mixture with an already low number of components. 

\section{Conclusions}
\label{sec:conclusions}
In this paper, a novel adaptive Gaussian mixture filter has been proposed. It is based on statistical linearization, which allows quantifying the induced linearization errors in terms of a linearization error covariance matrix. A criterion based on this covariance matrix is used for selecting Gaussian components for splitting, while the direction of the split is performed in direction of the eigenvalue with the strongest linearization errors. Compared to other splitting criteria, the proposed one reliably detects strong nonlinearities and keeps the number of splits on a low level. Furthermore, arbitrary approaches for statistical linearization can be employed.

%

%


\bibliographystyle{IEEEtran}

\end{document}